\documentclass[12pt]{article}
\usepackage[dvips]{epsfig}
\usepackage{harvard, color}
\usepackage{amsmath, amsthm, amssymb}
\usepackage{times}

\usepackage{xcolor}
\usepackage{multirow}
\usepackage{graphicx}
\usepackage{subfigure}
\usepackage{epstopdf}
\usepackage{caption}
\usepackage{listings}
\citationmode{abbr}
\usepackage{tikz}
\usetikzlibrary{shapes,snakes}
\usepackage{url}
\usepackage{hyperref}
\usepackage{booktabs}

\newtheorem{thm}{Theorem}

\newcommand{\bfX}{\mbox{\boldmath $X$}}
\newcommand{\bfD}{\mbox{\boldmath $D$}}

\def\0{{\bf 0}}

\def\squarebox#1{\hbox to #1{\hfill\vbox to #1{\vfill}}}

\def\sumIP1{\sum_{i=1, i\in P_1}^N}

\begin{document}
\title{\bf {\Large When Tukey meets Chauvenet: a new boxplot criterion for outlier detection}}

\author{Hongmei Lin$^1$, Riquan Zhang$^1$\thanks{Co-corresponding author} ~and Tiejun Tong$^{2}$\thanks{Co-corresponding author}\\
\\
{\small $^1$School of Statistics and Data Science, Shanghai University of International Business} \\
{\small  and Economics, Shanghai, China} \\
{\small $^2$Department of Mathematics, Hong Kong Baptist University, Hong Kong, China}}
\date{}
\maketitle

\begin{abstract}
\noindent
The box-and-whisker plot, introduced by \citeasnoun{Tukey1977bk}, is one of the most popular graphical methods in descriptive statistics.
On the other hand, however, Tukey's boxplot is free of sample size, yielding the so-called ``one-size-fits-all" fences for outlier detection.
Although improvements on the sample size adjusted boxplots do exist in the literature, most of them are either not easy to implement or lack justification.
As another common rule for outlier detection, Chauvenet's criterion uses the sample mean and standard derivation to perform the test,
but it is often sensitive to the included outliers and hence is not robust.
In this paper, by combining Tukey's boxplot and Chauvenet's criterion, we introduce a new boxplot, namely the Chauvenet-type boxplot, with the fence coefficient determined by an exact control of the outside rate per observation.
Our new outlier criterion not only maintains the simplicity of the boxplot from a practical perspective, but also serves as a robust Chauvenet's criterion.
Simulation study and a real data analysis on the civil service pay adjustment in Hong Kong demonstrate that the Chauvenet-type boxplot performs extremely well regardless of the sample size, and can therefore be highly recommended for practical use to replace both Tukey's boxplot and Chauvenet's criterion.
Lastly, to increase the visibility of the work, a user-friendly R package named `ChauBoxplot' has also been officially released on CRAN.

\end{abstract}

\newpage
\section{Introduction}

The box-and-whisker plot, introduced by \citeasnoun{Tukey1977bk}, is also known as Tukey's boxplot or, simply, the boxplot.
It is one of the most popular graphical methods in descriptive statistics, which provides a very efficient tool to visualize the distribution of the sample data.
The boxplot not only displays a box ranging from the first quartile ($Q_1$) to the third quartile ($Q_3$) of the data, but also provides two whiskers extending from the edge of the box to the lower and upper fences for outlier detection.
More specifically, by letting ${\rm IQR} = Q_3-Q_1$ be the interquartile range of the data, the lower and upper fences of the boxplot are given as ${\rm LF} = Q_1 - k\times {\rm IQR}$ and ${\rm UF} = Q_3 + k\times {\rm IQR}$, where $k>0$ is the fence coefficient.
\citeasnoun{Tukey1977bk} further referred to the fences with $k=1.5$ as the ``inner fences", and with $k=3$ as the ``outer fences".
The two visualized whiskers are then drawn from the smallest value within the inner fences to the first quartile, and from the third quartile to the largest value within the inner fences.
Lastly, the outliers that differ from the majority of the data will be plotted as individual points beyond the whiskers on the boxplot.
According to the degree of abnormality, \citeasnoun{Tukey1977bk} also declared the observations between the inner and outer fences as the ``outside" outliers,
and those beyond the outer fences as the ``far out" or ``extreme" outliers.
For more details on Tukey's boxplot and its variants for univariate data, one may refer to \citeasnoun{Frigge1989}, \citeasnoun{Schwertman2004}, \citeasnoun{Sim2005}, \citeasnoun{Wickham2011}, \citeasnoun{Walker2018}, \citeasnoun{Rodu2022}, and the references therein.

\vskip 6pt
Apart from the boxplot, the identification of outliers itself is another important issue in robust statistics and related areas.
Statistically, an outlier may be defined as an abnormal observation or contaminated data that does not follow the same distribution as the majority of the data.
If the outliers are not identified and further excluded from the data, they may have an unexpected impact on the subsequent data analysis, yielding a misleading or erroneous conclusion.
Because of this, there is a vast and growing literature on the identification and accommodation of suspected outliers since the 19th century.
To name a few, the first rigorous test was proposed by \citeasnoun{Peirce1852}, which is also known as Peirce's criterion.
Due to its complicated procedure, however, Peirce's criterion was soon superseded by an alternative called Chauvenet's criterion.
Given a sample of data from a normal distribution with $\bar X$ the sample mean and $S$ the sample standard deviation, \citeasnoun{Chauvenet1863} declared an observation as an outlier if its value is outside the interval $[\bar X - c_nS, \bar X+c_nS]$, where $c_n=\Phi^{-1}(1-0.25/n)$ is the upper $0.25/n$ quantile of the standard normal distribution.
Thanks to its simplicity and effectiveness, Chauvenet's criterion has been widely used in different fields including, but not limited to, astronomy, nuclear technology, geology, epidemiology, molecular biology, radiology and physical science \cite{Ross2003,Maples2018,Vermeesch2018}.

\vskip 6pt

It is noteworthy that there exist many other outlier testing procedures in the literature \cite{Andrews1974,Andrews1978,Beckman1983,Rosner1983,Davies1993,Fung1993,Atkinson1994,Carey1997,Bates2023,Bhattacharya2023},
as well as those being summarized in the monographs \cite{Hawkins1980bk,Barnett1994bk,Atkinson2000bk,Hawkins2006bk}.
To the best of our knowledge, Tukey's boxplot and Chauvenet's criterion are still among the most popular methods for outlier detection, mainly thanks to their simplicity as well as the reasonable accuracy \cite{Brant1990,Milton1999bk}.
Despite their popularity, each of these two methods has well-known disadvantages.
Specifically for Tukey's boxplot, we note that the fence coefficient $k$ is free of sample size, yielding the so-called ``one-size-fits-all" fences to label the outliers \cite{McGill1978,Sim2005}.
Although improvements on the sample size adjusted boxplots do exist in the literature, most of them are either not easy to implement \cite{Sim2005,Banerjee2007} or lack justification \cite{Barbato2011}.
For more details, see Section 2.
On the other hand, Chauvenet's criterion uses the sample mean and standard derivation to perform the test, and consequently, it is often very sensitive to the included outliers and, hence, may not provide a robust method for outlier detection.
For further illustration, see the toy example in Section 3.1.

\vskip 6pt
In this paper, by combining Tukey's boxplot and Chauvenet's criterion, we introduce a new boxplot, namely the {\it Chauvenet-type boxplot}, with the fence coefficient determined by an exact control of the outside rate per observation.
Our new outlier criterion not only maintains the simplicity of the boxplot from a practical perspective, but also serves as a robust Chauvenet's criterion.
Simulation study and a real data analysis on the civil service pay adjustment in Hong Kong also demonstrate that the Chauvenet-type boxplot performs extremely well regardless of the sample size, and can therefore be highly recommended for practical use to replace both Tukey's boxplot and Chauvenet's criterion.
The remainder of the paper is organized as follows.
In Section 2, we review the existing boxplots for outlier detection that incorporate the sample size into the construction of the lower and upper fences.
In Section 3, we first review Chauvenet's criterion and then propose our new boxplot that is a combination of Chauvenet's criterion and the classic boxplot for detecting outliers.
In Section 4, we conduct simulation study and real data analysis to compare the performance of the new and existing boxplots as well as Chauvenet's criterion.
In Section 5, we further extend the Chauvenet-type boxplot to handle non-normal data for outlier detection and demonstrate by simulations that our new criterion performs also very well.
Lastly, the paper is concluded in Section 6 with some discussion and future directions.

\vskip 12pt
\section{Existing boxplots for outlier detection}

As is known, Tukey's box-and-whisker plot is very appealing in descriptive statistics, mainly thanks to its simplicity and low sensitivity to outlier distortion.
On the other hand, however, the whiskers in Tukey's boxplot or some of its variants are free, or nearly free, of the sample size of the data so that they yield a {\it one-size-fits-all} rule for outlier detection.
In this section, we give a brief review of the major developments of boxplots that aim to incorporate the sample size into the fences.
For ease of presentation, we will assume the data is normally distributed in the next three sections, whereas the extensions to non-normal or skewed data will be presented in Section 5.

\vskip 6pt
Let $X_1,\dots,X_n$ be a sample of size $n$ from the normal distribution $N(\mu,\sigma^2)$, where $\mu$ is the mean and $\sigma^2>0$ is the variance.
Let also $Q_1$ and $Q_3$ be the first and third quartiles of the data, respectively.
To construct the boxplot, we define the lower and upper fences as
\begin{eqnarray}
{\rm LF}_n = Q_1 - k_n\times {\rm IQR} ~~~~~{\rm and}~~~~~ {\rm UF}_n = Q_3 + k_n\times {\rm IQR}, \label{fences}
\end{eqnarray}
where ${\rm IQR} = Q_3-Q_1$ is the interquartile range and $k_n$ is the fence coefficient depending on the sample size.
\citeasnoun{Hoaglin1986} termed the {\it some-outside rate per sample}, denoted by $\alpha$, as the probability that one or more observations in the normal sample will be wrongly classified as outliers. To control the some-outside rate, the outlier region is then set to satisfy
\begin{eqnarray}
P\left(\mbox{one~or~more~of~} X_1,\dots,X_n \in (-\infty, {\rm LF}_n)\cup ({\rm UF}_n, \infty)\right) = \alpha.  \label{some-outside}
\end{eqnarray}
In other words, if an observation $X$ is either smaller than ${\rm LF}_n$ or larger than ${\rm UF}_n$, then it will be labeled as an {\it $\alpha$-outlier} in the boxplot.
\citeasnoun{Hoaglin1986} further termed the {\it outlier labeling rule} to formally describe Tukey's outlier flagging procedure.

\vskip 6pt
Observing that the exact solution of $k_n$ satisfying Eq. (\ref{some-outside}) is rather complicated, \citeasnoun{Hoaglin1986} proposed a multistep approximation procedure to determine the choices of $k_n$ in the boxplot for samples with normal data.
One main disadvantage of their approximation procedure is that the statistical expertise and decision are needed in each step, which will unavoidably result in the propagation of approximation errors.
In view of this, \citeasnoun{Hoaglin1987} further applied the simulation to obtain the numerical values of $k_n$ for normal samples.
Specifically, by converting the boxplot outlier labeling rule to a formal outlier identification testing procedure, they found the value of $k_n$ to be close to 2.2 for $\alpha=0.05$ and $n$ up to 300.

\vskip 12pt
\subsection{Boxplots with exact some-outside rate}

For the boxplots with some-outside rate per sample, \citeasnoun{Sim2005} made some extra efforts to derive the exact solution of $k_n$ satisfying Eq. (\ref{some-outside}).
Let $X_{(1)} = \min\{X_1,\dots,X_n\}$ and $X_{(n)} = \max\{X_1,\dots,X_n\}$ be the minimum and maximum values of the sample.
Note that the event $\{\mbox{one~or~more~of~} X_1,\dots,X_n \in (-\infty, {\rm LF}_n)\cup ({\rm UF}_n, \infty)\}$ is the union of two disjoint compound events $A_1 = \{X_{(n)} > {\rm UF}_n\}$ and $A_2 = \{X_{(1)}< {\rm LF}_n,~ X_{(n)}\leq {\rm UF}_n\}$.
Under the exact control of the some-outside rate, the requirement on (\ref{some-outside}) implies that $P(A_1) + P(A_2) = \alpha$.
In addition, the standardization procedure to $X_1,\dots,X_n$ can be further applied so that the sample is free of $\mu$ and $\sigma$.
In this section, for simplicity of notation, we directly regard $X_1,\dots,X_n$ to be the standard normal data.
Finally, by the property that the order statistics of the sample form a Markov chain \cite{David2003bk}, an exact expression for $k_n$ satisfying Eq. (\ref{some-outside}) can be derived as
\begin{eqnarray*}
\int_{-\infty}^\infty \int_{x_1}^\infty \left\{1-I_{G_u(y_u)}(n-u,1) \left[1-I_{G_l(y_l)}(1,l-1) \right] \times f_{Q_1,Q_3}(x_1,x_3) \right\} dx_3dx_1 = \alpha,   \label{Eq.3}
\end{eqnarray*}
where $G_l(y_l) = \Phi(y_l)/\Phi(x_1)$ with $y_l = x_1-k_n(x_3-x_1)$, $G_u(y_u) = [\Phi(y_u)-\Phi(x_3)]/[1-\Phi(x_3)]$ with $y_u=x_3+k_n(x_3-x_1)$, $I_p(a,b)=\int_0^p t^{a-1}(1-t)^{b-1}dt/B(a,b)$ is the incomplete beta function, and $f_{Q_1,Q_3}(x_1,x_3)$ is the joint probability density function of $Q_1$ and $Q_3$.
For more details, see Eqs. (7) and (8) in \citeasnoun{Sim2005} together with the associated text.

\vskip 6pt
Next, to solve $k_n$ from the above question, \citeasnoun{Sim2005} further applied the multivariate quadrature rule to numerically approximate the double integral,
and then applied a direct search or Monte Carlo integration algorithm to search for the value of $k_n$ that yields an exact control of the some-outside rate per sample.
For practical use, they also provided some numerical values of $k_n$ in their Table 1 together with an approximation formula.
Taken the normal distribution with $\alpha=0.05$ and $n=4m+1$ with $m\in \{2,3,\dots,124\}$ as an example, the fence coefficient with exact some-outside rate (ER) is approximated as
\begin{eqnarray}
k_n^{\rm ER} \approx \exp\{4.01761 - 2.35363 \ln(n) + 0.64618 \ln^2(n) - 0.07893 \ln^3(n) + 0.00368 \ln^4(n)\}.  \label{k.ER}
\end{eqnarray}
For more numerical approximations with other values of $\alpha$ and $n$, one may refer to Table A.1 from their Appendix, in which the maximum absolute deviation $\delta$ between the true and approximated values of $k_n^{\rm ER}$ were also given.

\vskip 12pt
\subsection{Boxplots with tolerance limits}

As an alternative, \citeasnoun{Sim2005} further applied the tolerance interval, an important statistical tool in reliability and quality control, to construct the fences of the boxplot.
A tolerance interval is an interval determined from a random sample in such a way that one may have a specified $100\gamma$\% level of confidence that the interval will cover at least a specified $\beta$ proportion of the sampled population.
For more on tolerance intervals, one may refer to \citeasnoun{Guttman1970bk}, \citeasnoun{Patel1986}, and \citeasnoun{LiaoCT2004}.
For a certain distribution, let $f(x)$ be the probability density function and $F(x)$ be the cumulative distribution function.
Let also $L(\bfX) = L(X_1,\dots,X_n)$ and $U(\bfX) = U(X_1,\dots,X_n)$ be two statistics with $L(\bfX)<U(\bfX)$.
Then with the probability coverage of the random interval $W = \int_{\scriptsize L(\bfX)}^{\scriptsize U(\bfX)} f(x)dx = F(U(\bfX)) - F(L(\bfX))$,
the interval $[L(\bfX),U(\bfX)]$ defines a two-sided $(\gamma,\beta)$ tolerance interval if $P(W\geq \beta)=\gamma$, where $\gamma$ and $\beta$  are two specified probability values.

\vskip 6pt
As a good example, the tolerance interval with $L(\bfX)=\bar X - kS$ and $U(\bfX)=\bar X + kS$ has been widely used in quality control \cite{Guenther1977bk}, where $\bar X$ is the sample mean and $S$ is the sample standard deviation.
Now with the boxplot, \citeasnoun{Sim2005} constructed a $(\gamma,\beta)$ tolerance interval with $L(\bfX)$ and $U(\bfX)$ being the lower and upper fences in (\ref{fences}), respectively.
Observations that fall outside the tolerance limits are labeled as outliers.
Then by letting $\beta=1-\alpha$, the value of $k_n$ is defined to satisfy $P(P(\mbox{all~of~} X_1,\dots,X_n \in ({\rm LF}_n, {\rm UF}_n)\geq 1-\alpha)) = \gamma$, or equivalently,
\begin{eqnarray*}
P\left(\int_{{\rm LF}_n}^{{\rm UF}_n} f(x)dx \geq 1-\alpha_n\right) = \gamma,
\end{eqnarray*}
where $\alpha_n = 1 - (1-\alpha)^{1/n}$.
Moreover, by the symmetry of the normal distribution, the value of $k_n$ can be determined from the equation $P(P(X<{\rm LF}_n)\leq \alpha_n/2)=\gamma$.
Finally, numerical approximation using the quadrature rule needs to be implemented for computing the integral, followed by a direct search algorithm or Monte Carlo integration algorithm to search for the value of $k_n$.
For practical use, the authors also provided the numerical values of $k_n$ in their Table 2 for different combinations of $(\alpha, \gamma, n)$, together with the approximation formulas in Table A.2.
Taken the normal distribution with $\alpha=0.05$, $\gamma=0.9$ and $n=4m+1$ with $m\in \{2,3,\dots,124\}$ as an example, the fence coefficient with tolerance limits (TL) is approximated as
\begin{eqnarray}
k_n^{\rm TL} \approx \exp\{4.45171 - 2.44501 \ln(n) + 0.64990 \ln^2(n) - 0.07851 \ln^3(n) + 0.00365 \ln^4(n)\}.  \label{k.TL}
\end{eqnarray}

\vskip 12pt
\subsection{Boxplots with asymptotic fences}

In view of the complexity in deriving the exact expression of $k_n$ satisfying Eq. (\ref{some-outside}), \citeasnoun{Iglewicz2001} and \citeasnoun{Banerjee2007} proposed an alternative solution based on the asymptotics.
Specifically, they first derived the limiting solution of $k_n$ from Eq. (\ref{some-outside}) when $n$ goes to infinity, and then regress the fitted values back to smaller sample sizes.
By the symmetry of the normal distribution and the fact that $\{\mbox{one~or~more~of~} X_1,\dots,X_n \in (-\infty, {\rm LF}_n)\cup ({\rm UF}_n, \infty)\} = \{X_{(1)}<{\rm LF}_n~{\rm or}~ X_{(n)}>{\rm UF}_n\}$, \citeasnoun{Banerjee2007} thus simplified the solution to Eq. (\ref{some-outside}), approximately, as the solution to the equation of
\begin{eqnarray*}
P\left(X_{(n)} \leq Q_3 + k_n \times {\rm IQR} \right) = 1-\alpha/2.  \label{X.n}
\end{eqnarray*}
Finally, since $P(X_{(n)}\leq x) = [P(X_1\leq x)]^n$, it suffices to solve $P(X_1 \leq Q_3 + k_n \times {\rm IQR}) = (1-\alpha/2)^{1/n}$ for the value of $k_n$.

\vskip 6pt
For normal data, as mentioned in Section 2.1, the solution of $k_n$ to the above question does not depend on $\mu$ and $\sigma$.
Thus for simplicity, we assume that the data is sampled from $N(0,1)$ with $\Phi(x)$ being the cumulative distribution function.
When $n$ is sufficiently large, by replacing $Q_1$ by $\Phi^{-1}(0.25)$ and $Q_3$ by $\Phi^{-1}(0.75)$, it leads to the limiting solution as
$k_n = [\Phi^{-1}(1-\alpha/2)^{1/n}) - \Phi^{-1}(0.75)]/(\Phi^{-1}(0.75) - \Phi^{-1}(0.25)) = [\Phi^{-1}(1-\alpha/2)^{1/n}) - 0.6745]/1.349$.
Finally, on the basis of a simulation study for smaller sample sizes, \citeasnoun{Banerjee2007} recommended their fence coefficient with asymptotic fences (AF) as
\begin{eqnarray}
k_n^{\rm AF} \approx a_n \times {\Phi^{-1}\left((1-\alpha/2)^{1/n}\right) - 0.6745 \over 1.349},   \label{k.AF}
\end{eqnarray}
where $a_n$ is the smoothed adjustment value with $a_n = 1 + 8.9764n^{-1} - 126.6262n^{-2} + 1531.7064n^{-3} - 10729.3439n^{-4}$ for $n<2000$, and $a_n=1$ for $n\geq 2000$.

\vskip 12pt
\subsection{Boxplots with empirical considerations}

Recall that Tukey's outlier flagging procedure takes a constant $k$ when constructing the lower and upper fences, yielding a one-size-fits-all boxplot for labeling the potential outliers.
To take into account the sample size, unlike those complicated procedures in the previous sections that control either the some-outside rate or the tolerance limit,
\citeasnoun{Barbato2009bk} suggested a modified IQR method based on some empirical considerations (EC).
Specifically, by modifying IQR with IQR$[1+0.1\ln(n/10)]$, their fence coefficient associated with $k=1.5$ is given as
\begin{eqnarray}
k_n^{\rm EC} = 1.5 \times \left[1+0.1 \ln(n/10)\right]. \label{k.EC}
\end{eqnarray}
Note that $k_n^{\rm EC}$ is monotonically increasing with $n$, and so is the interval length consisting of the two fences, in such a way that the dependence on the sample size is introduced into the boxplot.
In addition, as stated by \citeasnoun{Barbato2011}, their boxplot outlier labeling rule retains its primary advantages, including simplicity and the lack of need for iteration, thanks to the robust estimate of spread achieved by excluding both tails.

\vskip 6pt
We are not aware of any newer developments for constructing the sample size adjusted boxplots with univariate normal data, expect for those with non-normal or skewed data for which we will discuss later in Section 6.
It is also noteworthy that, during the last decade, researchers have turned their main attention to other types of boxplots including, for example, boxplots for functional data \cite{Hyndman2010,SunY2011,DaiW2018,QuZ2022}, boxplots for circular data \cite{Abuzaid2012,Buttarazzi2018}, boxplots for contours, curves and paths \cite{Whitaker2013,Mirzargar2014,Raj2017}, and boxplots for large data \cite{Hofmann2017}.

\vskip 24pt
\section{Chauvenet-type boxplot}

As reviewed in Section 2, the whiskers in Tukey's boxplot are free of the sample size of the data, yielding the so-called ``one-size-fits-all" criterion for outlier detection.
Although improvements on the sample size adjusted boxplots do exist in the literature, most of them are either not easy to implement or lack justification.
As another example for illustration, on the basis of \citeasnoun{Hoaglin1986} and \citeasnoun{Hoaglin1987}'s boxplot outlier labeling rules,
\citeasnoun{Frigge1989} further conducted a survey on popular software packages to standardize the fence coefficient selection for boxplot construction.
It is noted, however, that they finally recommended to use \citeasnoun{Tukey1977bk}'s fence constants of $1.5$ or $3$ as the default value in the boxplot.
This, from another perspective, shows that the sample size adjusted whiskers in their boxplot papers in 1986 and 1987 may not be very practical for outlier detection.

\vskip 12pt
\subsection{Chauvenet's criterion}

Apart from the boxplot that uses the lower and upper quartiles, another common approach is to construct the outlier regions based on the mean ($\bar X$) and standard deviation ($S$) of the sample data.
Among this category, the {\it $3$-sigma limits} are the best known in the quality control chart \cite{Pukelsheim1994,Wheeler2010bk}, by which a data point will be labeled as a suspected outlier if outside the interval $[\bar X - 3S, \bar X+3S]$.
As can be seen, this method is relatively simple to implement, and in addition, the constant $3$ is not a must as the multiplier of the standard deviation.
For instance, $[\bar X - 2S, \bar X+2S]$, $[\bar X - 2.5S, \bar X+2.5S]$ and $[\bar X - 4S, \bar X+4S]$ are also occasionally used, depending on whether a more relaxed or strict control is desired.
More interestingly, with {\it a constant time of sigma} clipping, we note that the 3-sigma limits work in a similar way as Tukey's whiskers with {\it a constant time} of ${\rm IQR}$ as the length.
Taking the normal data as an example, Tukey's outlier labeling rule with $k_n=1.5$ or $k_n=1.72$ is asymptotically equivalent to the $2.62$-sigma or $3$-sigma clipping, respectively \cite{Morales2021}.

\vskip 6pt
To incorporate the sample size into the sigma clipping rules, the first rigorous test using the probability theory was proposed by Peirce in 1852, which is also known as Peirce's criterion \cite{Peirce1852,Gould1855,Peirce1878}.
Peirce's test is based on what are now called the $Z$-scores, by which any observations with $|Z|>c$ will be rejected, where $c$ depends on both the sample size and the number of suspected outliers.
Due to the complicated form of $c$, however, Peirce's test is not easy to implement which greatly hinders its popularity for practical use.
And in fact, it was soon superseded by an alternative called {\bf Chauvenet's criterion}, which identifies a data point as an outlier if the probability of obtaining a value at least as extreme as the observed one is less than $0.5/n$, where $n$ is the size of the sample.
Statistically, by letting $X_1,\dots,X_n$ be a random sample from $N(\mu,\sigma^2)$ with $\bar X$ as the sample mean and $S$ as the sample standard deviation, \citeasnoun{Chauvenet1863} first computed the deviation of each data point from the mean as $D_i = |X_i - \bar X|/S$, where $i=1,\dots,n$.
He then set $c_n=\Phi^{-1}(1-0.25/n)$ as the threshold, and declared each data point with deviation $D_i$ greater than $c_n$ as a suspected outlier.
Or equivalently, if we present the criterion using the sigma clipping, then a data point will be labeled as an outlier if outside the interval
\begin{eqnarray}
[\bar X - c_nS, \bar X+c_nS].  \label{Chauvenet.interval}
\end{eqnarray}
To summarize, Chauvenet's criterion will reject, on average, half an observation of genuine data from the normal distribution regardless of the sample size.

\vskip 6pt
As mentioned in Section 1, Chauvenet's criterion has been widely used in many different fields thanks to its simplicity and effectiveness \cite{Ross2003}, as well as its usage spread across government laboratories, industry, and universities.
As a more recent example, Chauvenet's criterion and its modified version have been adopted for outlier detection by the very popular toolbox, {\it IsoplotR}, for geochronology \cite{Vermeesch2018}, which has been cited more than 2500 citations in Google Scholar as of December 2024.
Despite its popularity, Chauvenet's criterion has drawbacks.
An obvious limitation is that it can be very sensitive to the included outliers and, hence, may not provide a robust method for outlier detection.

\vskip 6pt
To illustrate why Chauvenet's criterion is lack of robustness, we also consider a toy example with $n=9$ observations, where $X_1,\dots,X_7$ are a random sample of size $7$ from the standard normal distribution and $X_8=X_9=100$ are two contaminated data points. For ease of presentation, we further sort the observations in ascending order so that the whole data set is $\bfX = \{-1.938, -1.177, -0.854, -0.353, 0.890, 0.916, 1.741, 100, 100\}$.
Then with the sample mean $\bar X = 22.136$ and standard deviation $S = 44.160$, the deviations of the data are given as
\begin{eqnarray*}
\bfD = {|\bfX - \bar X| \over S} = \{0.545, 0.528, 0.521, 0.509, 0.481, 0.481, 0.462, 1.763, 1.763\}.
\end{eqnarray*}
Finally, since the maximum deviation $D_{\rm max}=1.763$ is less than the threshold $c_9=\Phi^{-1}(1-0.25/9) = 1.915$, Chauvenet's criterion will hence declare no data point as suspected outliers, although the last two are clearly contaminated.

\vskip 12pt
\subsection{A new boxplot criterion for outlier detection}

As seen from the above toy example, Chauvenet's criterion, and more generally the sigma clipping rules, suffer from the included outliers via the contaminated sample mean and standard deviation.
To improve the robustness, better estimates of the mean and standard deviation ought to be considered, e.g. using the quantile information as given in Tukey's boxplot.
On the other hand, as reviewed in Section 2, Tukey's outlier labeling rule and its existing variants are themselves not perfect in the way of incorporating the sample size.
To further improve boxplots for outlier detection, effective strategies involving the sample size adjusted sigma clipping can be incorporated.

\vskip 6pt
Inspired by this, we propose to take advantage of both Tukey's boxplot and Chauvenet's criterion.
More specifically, through the fusion of these two classic technologies, we are able to introduce a new boxplot criterion for more accurately and more robustly labeling the outliers of the data.
By formula (\ref{fences}), recall that the lower and upper fences in Tukey's boxplot are ${\rm LF}_n = Q_1 - k_n\times {\rm IQR}$ and ${\rm UF}_n = Q_3 + k_n\times {\rm IQR}$.
Now to apply Chauvenet's criterion to determine the $k_n$ value, we first transform the boxplot fences from the two quartiles, $Q_1$ and $Q_3$, back to the sample mean and standard deviation.
For normal data, by \citeasnoun{Tukey1977bk} we can apply the midhinge $(Q_1+Q_3)/2$ to estimate the sample mean, yielding $Q_1+Q_3 \approx 2\bar X$;
Further by \citeasnoun{higgins2019cochrane}, we apply $(Q_3-Q_1)/1.35$ to estimate the sample standard deviation, yielding $Q_3-Q_1 \approx 1.35S$.
Solving the above two equations, we have $Q_1 \approx \bar X - 0.675S$ and $Q_3 \approx \bar X + 0.675S$.
Finally, by plugging them into (\ref{fences}), the new lower and upper fences represented by the sample mean and standard deviation can be expressed as ${\rm LF}_n = \bar X - (1.35k_n+0.675)S$ and ${\rm UF}_n = \bar X + (1.35k_n+0.675)S$.

\vskip 6pt
Now to fuse Tukey's boxplot with Chauvenet's criterion, we apply the upper $0.25/n$ quantile of the standard normal distribution, i.e. $c_n=\Phi^{-1}(1-0.25/n)$, as the threshold for labeling anomalous observations.
This leads to $1.35k_n+0.675=\Phi^{-1}(1-0.25/n)$, or equivalently, $k_n= \Phi^{-1}(1-0.25/n)/1.35-0.5$.
Taken together, the lower and upper fences of our new boxplot criterion for outlier detection are, respectively,
\begin{eqnarray}
{\rm LF}_n^{\rm Chau} = Q_1 - k_n^{\rm Chau} \times {\rm IQR} ~~~~~{\rm and}~~~~~{\rm UF}_n^{\rm Chau} = Q_3 + k_n^{\rm Chau} \times {\rm IQR},  \label{Chau.fence}
\end{eqnarray}
where the fence coefficient associated with Chauvenet's criterion is
\begin{eqnarray}
k_n^{\rm Chau} = {\Phi^{-1}(1-0.25/n) \over 1.35} - 0.5.  \label{Chau.coefficient}
\end{eqnarray}
We refer to the above boxplot with new whiskers as the {\it Chauvenet-type boxplot}.
With the new fence coefficient $k_n^{\rm Chau}$ in (\ref{Chau.coefficient}), the Chauvenet-type boxplot not only provides an exact control of the outside rate per observation, but also well maintains the simplicity of the boxplot from the perspective of practice.

\vskip 6pt
Lastly, to demonstrate that our new boxplot criterion is more robust for outlier detection than Chauvenet's criterion, we revisit the toy example in Section 3.1.
We have $Q_1=-0.854$ and $Q_3=1.741$, which are indeed the 3rd and 7th observations respectively.
Further by (\ref{Chau.fence}) and (\ref{Chau.coefficient}), we have $k_n^{\rm Chau} = 0.918$, and consequently, $[{\rm LF}_n^{\rm Chau}, {\rm UF}_n^{\rm Chau}] = [-3.237, 4.124]$.
Finally, observing that the two contaminated data points with value 100 are far beyond the upper fence at 4.124, our new boxplot criterion for outlier detection will thus declare them as extreme outliers.
On the other hand, the lower and upper limits using Chauvenet's criterion are given as $\bar X \pm 1.915S = [-62.430, 106.702]$, providing a very wide interval that contains all 9 observations, and hence fail to declare any anomalous data points.
To conclude, it is evident that our new boxplot criterion has provided a robust version of Chauvenet's criterion, in such a way that the lower and upper fences will no longer be sensitive to the extreme values when they present.

\vskip 12pt
\subsection{Further exploration}

When the data is normal, the following theorem establishes an asymptotic equivalence between our new boxplot criterion and Chauvenet's criterion for outlier detection, with the proof being relegated to Section S1 of the online Supplementary Material.

\vskip 6pt
\begin{thm} \label{theorem1}
For normal data, the outlier region of the Chauvenet-type boxplot is asymptotically equivalent to that of Chauvenet's criterion. More specifically by (\ref{Chauvenet.interval}) and (\ref{Chau.fence}), we have
\begin{align}
\bar X - c_nS &\simeq Q_1 - k_n^{\rm Chau} \times {\rm IQR}, \label{AE1}\\
\bar X + c_nS &\simeq Q_3 + k_n^{\rm Chau} \times {\rm IQR}, \label{AE2}
\end{align}
where $\simeq$ denotes the asymptotic equivalence such that if $a_n\simeq b_n$, then $a_n/b_n\to 1$ as $n\to\infty$.
\end{thm}

\vskip 6pt
By Theorem \ref{theorem1}, the Chauvenet-type boxplot provides a comparable performance to Chauvenet's criterion when the normal data is not contaminated.
But when the data contains outliers, both $\bar X$ and $S$ may be dramatically affected so that the lower and upper thresholds of Chauvenet's criterion in (\ref{Chauvenet.interval}) tends be wider, or much wider, than is expected.
In contrast, our new boxplot criterion will be statistically resistant to the presence of outliers.
Specifically by the definition in \citeasnoun{Huber2004bk}, the breakdown point of our new boxplot criterion will be as large as $25\%$ of the whole data.
In other words, when the number of outliers do not exceed $25\%$, the lower and upper fences in (\ref{Chau.fence}) for our new boxplot criterion will remain unchanged.
Taking the toy example for illustration, with two contaminated data being 100, the percentage of the outliers is $2/9=22.2\%<25\%$ and so they will not affect the outlier region.
This explains why our new boxplot criterion can successfully detect the two anomalous data points, whereas Chauvenet's criterion fails to do so due to the very wide interval (\ref{Chauvenet.interval}) caused by the contaminated sample mean and standard deviation.

\begin{figure}
\center{
\includegraphics[width=14cm, angle=0]{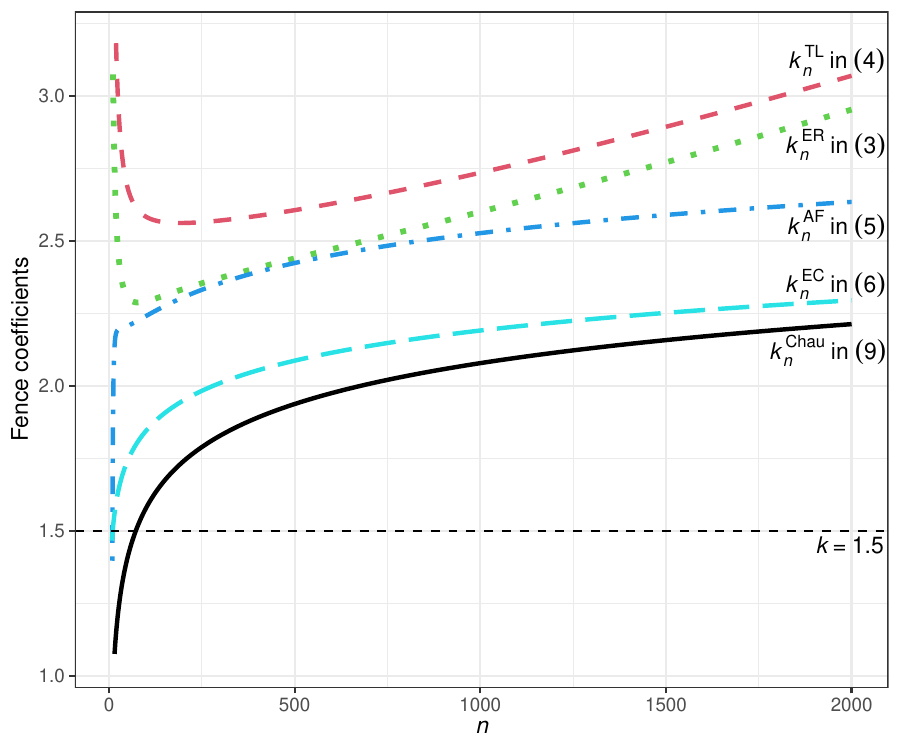}
\caption{The sample size adjusted fence coefficients for the new and existing boxplots, where $n$ ranges from 8 to 2000.
The dotted line represents the fence coefficient $k_n^{\rm ER}$ in (\ref{k.ER}),
the dashed line represents the fence coefficient $k_n^{\rm TL}$ in (\ref{k.TL}),
the dash-dotted line represents the fence coefficient $k_n^{\rm AF}$ in (\ref{k.AF}),
the long-dashed line represents the fence coefficient $k_n^{\rm EC}$ in (\ref{k.EC}),
and the solid line represents our new fence coefficient $k_n^{\rm Chau}$ in (\ref{Chau.coefficient}).
Lastly, the fence constant of $k=1.5$ in Tukey's boxplot is also given for reference.} \label{figure1}
}
\end{figure}

\vskip 6pt
Next, to explore the finite sample performance of the Chauvenet-type boxplot, we further plot the values of $k_n^{\rm Chau}$ in Figure \ref{figure1} with the sample size ranging from 8 to 2000.
And for comparison, the existing fence coefficients reviewed in Section 2 are also ploted, including the fence coefficient $k_n^{\rm ER}$ in (\ref{k.ER}), the fence coefficient $k_n^{\rm TL}$ in (\ref{k.TL}), the fence coefficient $k_n^{\rm AF}$ in (\ref{k.AF}), and the fence coefficient $k_n^{\rm EC}$ in (\ref{k.EC}).
By Figure \ref{figure1}, it is evident that the existing sample size adjusted methods all provide a longer whisker, and hence a more stringent criterion, than our new boxplot criterion for labeling the suspected outliers.
We also note that our new fence coefficient is much closer to Tukey's fence constant with $k=1.5$, and in particular, the two fence coefficients will be the same when $n=72$.
In other words, with the sample size less than 72, we recommend a whisker even shorter than that in Tukey's boxplot for outlier detection, in a way to more effectively increase the detection power for outliers.
In contrast, we observe that the two fence coefficients in \citeasnoun{Sim2005}, including $k_n^{\rm ER}$ in (\ref{k.ER}) and $k_n^{\rm TL}$ in (\ref{k.TL}), both display a decreasing pattern when $n$ is small which may not be reasonable.
While for the fence coefficient $k_n^{\rm AF}$ in (\ref{k.AF}), it has a rapidly increasing trend when $n$ is less than 20, thus yielding a less-smooth function of the sample size.

\vskip 12pt
\subsection{R package}

To increase the visibility of the work, a user-friendly R package named `ChauBoxplot' has also been officially released on CRAN.
To install and load the package, one may enter the following commands in the R console:

$>$ \texttt{install.packages("ChauBoxplot")}

$>$ \texttt{library(ChauBoxplot)}

\noindent
The new package consists of two main functions, {\it chau\underline{~~}boxplot()} and {\it geom\underline{~~}chau\underline{~~}boxplot()}, for graphically drawing the Chauvenet-type boxplot.
For the first function, it can be operated the same way as {\it boxplot()} in `base R', except that the new fence coefficient $k_n^{\rm Chau}$ in (\ref{Chau.coefficient}) is now adopted rather than the default $k=1.5$.
For the second function, it can also be operated the same way as {\it geom\underline{~~}boxplot()} in `ggplot2'.
Lastly, for reference, the source files and example codes are freely accessible on GitHub at \url{https://github.com/tiejuntong/ChauBoxplot}.

\vskip 24pt
\section{Comparison of the boxplots}

We compare the performance of the new and existing boxplots for outlier detection using normal data in Section 4.1, non-normal data in Section 4.2, and real data in Section 4.3.
Note that the sample size adjusted whiskers, as reviewed in Section 2, are seldom used in practice due to the difficulty to implement or lack of justification.
Thus to save space, we will only present the comparison results for the Chauvenet-type boxplot with the fence coefficient $k_n^{\rm Chau}$ in (\ref{Chau.coefficient})
and the classic Tukey's boxplot with the fence coefficient $k=1.5$.
We will demonstrate that our new boxplot criterion associated with the Chauvenet-type boxplot is not only simple and practical, but also more accurate and robust in labeling outliers.

\vskip 12pt
\subsection{Normal data}

\begin{figure}
\center{
\includegraphics[width=16.5cm, angle=0]{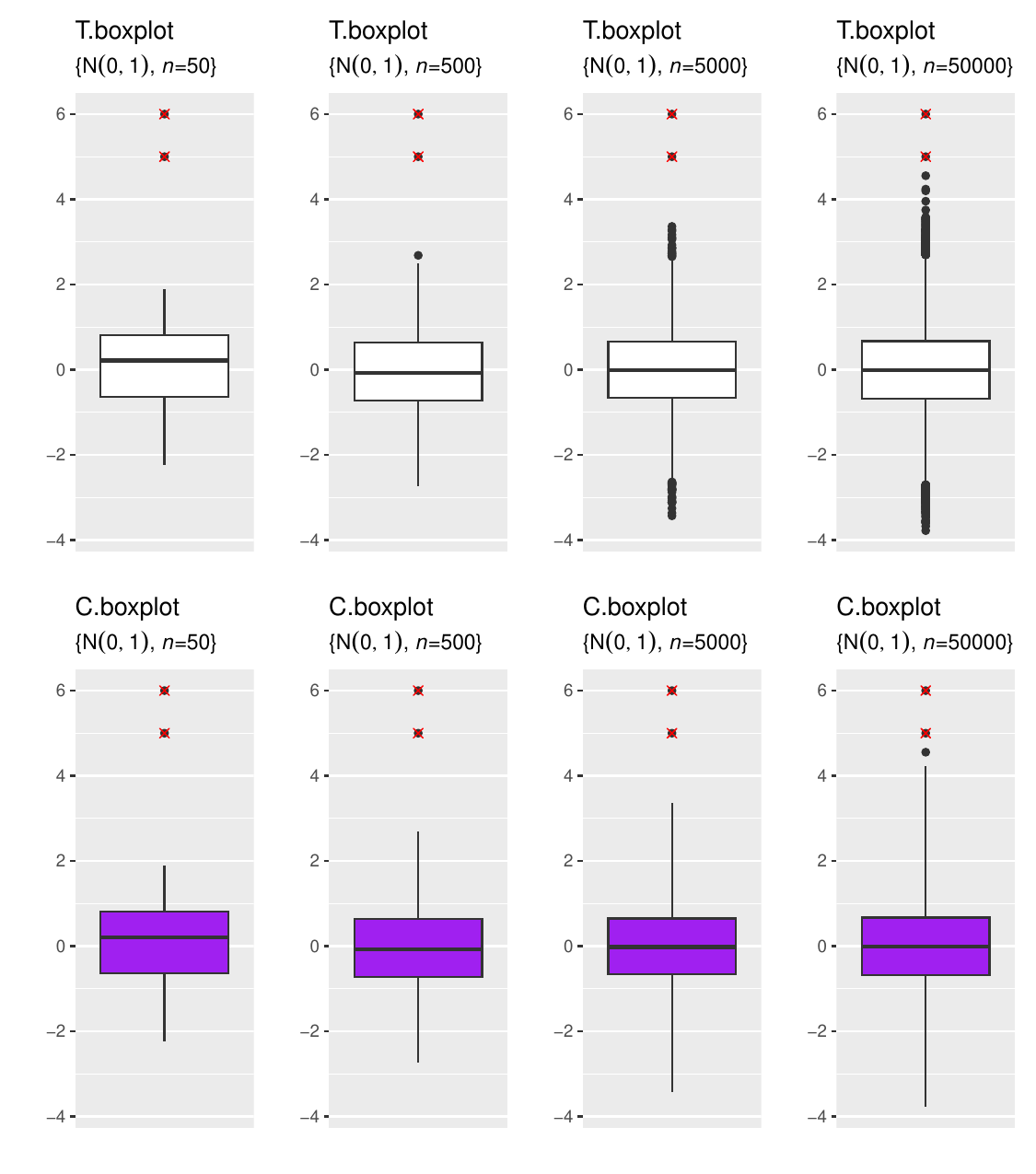}
\caption{Outlier detection for normal data using Tukey's boxplot (T.boxplot) and the Chauvenet-type boxplot (C.boxplot), where the data is generated from the standard normal distribution with two contaminated data as 5 and 6. With {\it set.seed(1863)} in R, Tukey's boxplot labels a total of 2, 3, 48 and 357 suspected outliers, and the Chauvenet-type boxplot labels a total of 2, 2, 2 and 3 suspected outliers, for the simulated data with $n=50$, 500, 5000 and 50000, respectively.
Lastly, the double glyph, consisting of a cross and a dot, indicates that the observation is both contaminated data and labeled as an outlier.} \label{figure2}
}
\end{figure}

Given a total sample size of $n$, we generate the first $n-2$ observations as a random sample from the standard normal distribution, and designate the last two observations as contaminated data with $X_{n-1}=5$ and $X_n=6$.
Then with $n=50$, 500, 5000 and 50000, we plot the simulated data in Figure \ref{figure2} using Tukey's boxplot (T.boxplot) and the Chauvenet-type boxplot (C.boxplot).
From the figure, we note that Tukey's boxplot will label more and more genuine data mistakenly as outliers when the sample size becomes larger.
As an example, with {\it set.seed(1863)} in R, Tukey's boxplot will label a total of 3, 48 and 357 suspected outliers when $n=500$, 5000 and 50000, respectively.
By excluding the two contaminated data, the remaining 1, 46 and 355 outliers are indeed genuine data from the standard normal distribution and so are wrongly labeled,
which coincides with the well-known statement that ``{\it if the data is sampled from the normal distribution, then about 0.7\% of them will lie outside the two fences of the boxplot}" \cite{Hubert2008}.
While for the Chauvenet-type boxplot, it labels a total of 2, 2, 2 and 3 suspected outliers for the simulated data with 4 different sample sizes, respectively.
If we further exclude the two contaminated data, it then only labels 0, 0, 0 and 1 genuine data as outliers by mistake for each boxplot.
This coincides with Chauvenet's criterion that it will reject, on average, half an observation of genuine data from the normal distribution, regardless of how large the sample size is.

\vskip 12pt
\subsection{Non-normal data}

\begin{figure}
\center{
\includegraphics[width=16.5cm, angle=0]{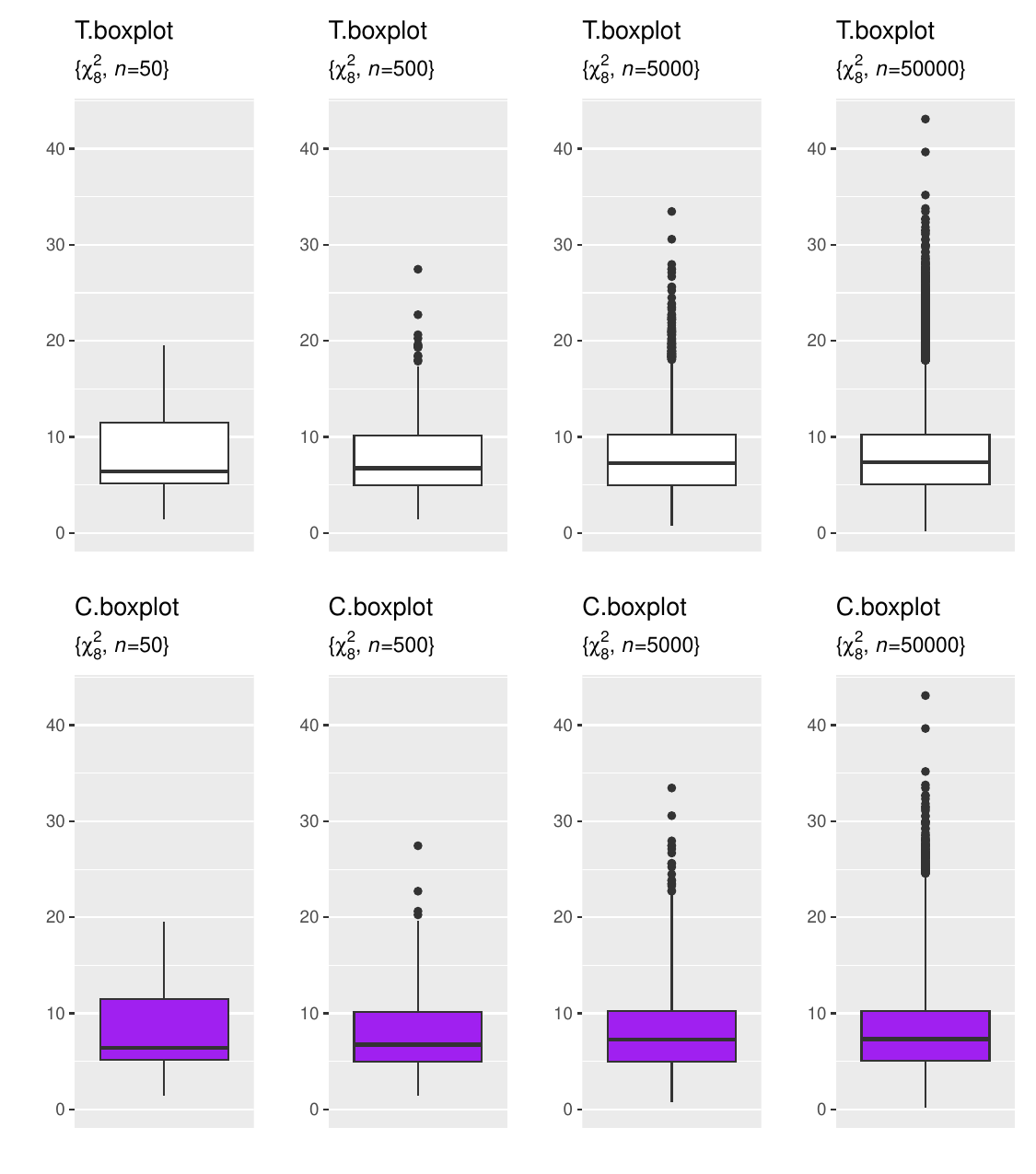}
\caption{Outlier detection for skewed data using Tukey's boxplot (T.boxplot) and the Chauvenet-type boxplot (C.boxplot), where the data is generated from the chi-square distribution with 8 degrees of freedom. With {\it set.seed(1863)} in R, Tukey's boxplot labels a total of 0, 13, 102 and 1102 suspected outliers, and the Chauvenet-type boxplot labels a total of 0, 4, 18 and 106 suspected outliers, for the simulated data with $n=50$, 500, 5000 and 50000, respectively.} \label{figure3}
}
\end{figure}

\begin{figure}
\center{
\includegraphics[width=16.5cm, angle=0]{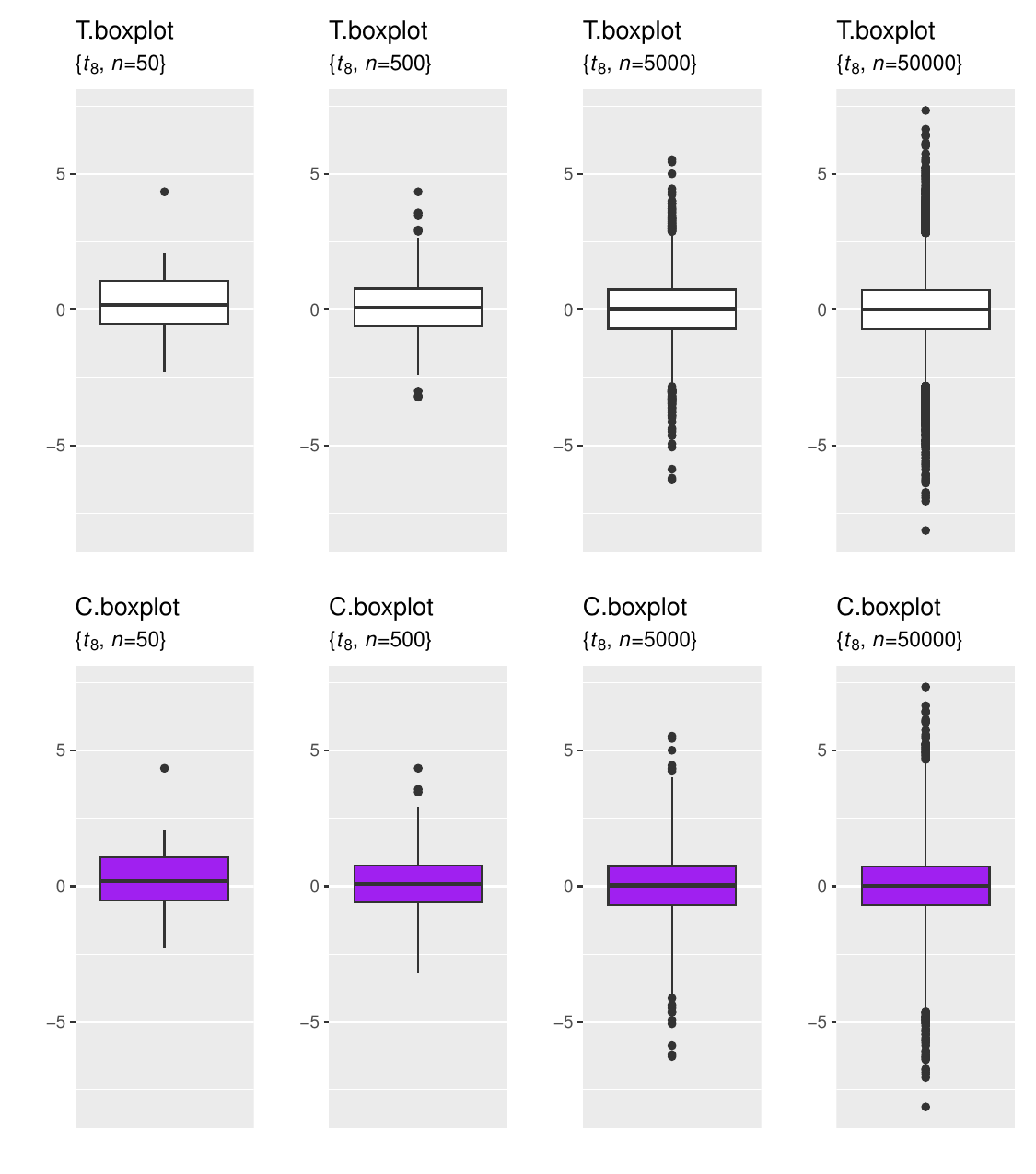}
\caption{Outlier detection for heavy tailed data using Tukey's boxplot (T.boxplot) and the Chauvenet-type boxplot (C.boxplot), where the data is generated from Student's $t$ distribution with 8 degrees of freedom. With {\it set.seed(1863)} in R, Tukey's boxplot labels a total of 1, 8, 117 and 1104 suspected outliers, and the Chauvenet-type boxplot labels a total of 1, 3, 18 and 90 suspected outliers, for the simulated data with $n=50$, 500, 5000 and 50000, respectively.} \label{figure4}
}
\end{figure}

For non-normal data, we first consider two common distributions.
One is the chi-square distribution, which is right skewed, and the other is Student's $t$ distribution, which is symmetric but heavy tailed.
We also consider 8 and 30 to represent the small and large degrees of freedom associated with each distribution, but to save space, we only present the simulation results for 8 degrees of freedom in the main text.
Then with $n=50$, 500, 5000 and 50000, we generate sample data from the two distributions and then plot them in Figures \ref{figure3} and \ref{figure4}, respectively, using Tukey's boxplot (T.boxplot) and the Chauvenet-type boxplot (C.boxplot).
As is seen, there is no contaminated data involved in this simulation; in other words, this study does not suffer from any real outliers.
Nevertheless, the two figures show that both Tukey's boxplot and the Chauvenet-type boxplot have wrongly labelled many outliers, especially when the sample size is large.
Notably, a similar pattern has also been observed from other skewed or heavy tailed distributions, including, but not limited to, the beta, exponential, gamma, and log-normal distributions.
Fore more details, please refer to the boxplots in the online Supplementary Material.

\vskip 6pt
For a closer comparison, we note that the two boxplots perform similarly when $n=50$, mainly because the fence coefficient by Chauvenet's criterion, $k_{50}^{\rm Chau}=1.41$, is very close to Tukey's coefficient at $k=1.5$.
When the sample size becomes larger, both of the boxplots will increasingly label more genuine data as outliers, but our new boxplot criterion is relatively much better.
Taking {\it set.seed(1863)} again in R, for the $\chi_8^2$ distribution with $n=500$, 5000 and 50000 respectively, Tukey's boxplot labels a total of 13, 102 and 1102 suspected outliers, whereas our new boxplot only labels 4, 18 and 106 suspected outliers.
And for the $t_8$ distribution with $n=500$, 5000 and 50000 respectively, Tukey's boxplot labels a total of 8, 117 and 1104 suspected outliers, whereas our new boxplot only labels 3, 18 and 90 suspected outliers.
To conclude, although not perfect, our newly developed Chauvenet-type boxplot has the capacity of labeling less ``false positive" outliers thanks to the sample size adjusted fences, especially for large sample sizes.

\vskip 24pt
\subsection{Real data}

For real data, we consider the civil service pay adjustment in Hong Kong, where the data can be downloaded from \url{https://www.csb.gov.hk/english/admin/pay/55.html} for the annual tax year which runs from April 1 to March 31 of the following year.
In the twenty-four tax years since 2001, civil servants in Hong Kong had their pay frozen or cut a total of eight times.
In particular, as a consequence of the SARS outbreak, they suffered from two immediate pay cuts in 2003-2004 and 2004-2005, followed by two pay freezes in 2005-2006 and 2006-2007.
Since 2007, the economic situation of Hong Kong has improved significantly, but still encountered two adverse events thereafter.
The first was the global financial crisis between mid 2007 and early 2009, which resulted in a pay cut of 5.38\% for the upper-level staff and a pay freeze for the lower and mid-level staff in the tax year of 2009-2010.
And the second was the economic turmoil following the COVID-19 pandemic, which also prompted a pay freeze for two consecutive years in 2020-2021 and 2021-2022.

\vskip 12pt
\begin{table}[h]
\small\sf\centering
\caption{The annual rates of pay adjustment in Hong Kong for the lower and mid-level staff (Junior civil servants) and the upper-level staff (Senior civil servants) since 2007.
Among the 18 tax years, junior civil servants had suffered a total of 3 pay freezes in 2009-2010, 2020-2021 and 2021-2022.
And senior civil servants had suffered a pay cut of 5.38\% in 2009-2010, followed by 2 pay freezes in 2020-2021 and 2021-2022.} \label{HK_salary}
\begin{tabular}{c|cc}
\toprule
~~~~Tax year~~~~ & Junior civil servants & Senior civil servants  \\
\midrule
2024-2025 &	3.00\%	&  3.00\% \\
2023-2024 &	4.65\%	&  2.87\% \\
2022-2023 &	2.50\%	&  2.50\% \\
2021-2022 &	0.00\%	&  0.00\% \\
2020-2021 &	0.00\%	&  0.00\% \\
2019-2020 &	5.26\%	&  4.75\% \\
2018-2019 &	4.51\%	&  4.06\% \\
2017-2018 &	2.94\%	&  1.88\% \\
2016-2017 &	4.68\%	&  4.19\% \\
2015-2016 &	4.62\%	&  3.96\% \\
2014-2015 &	4.71\%	&  5.96\% \\
2013-2014 &	3.92\%	&  2.55\% \\
2012-2013 &	5.80\%	&  5.26\% \\
2011-2012 &	6.16\%	&  7.24\% \\
2010-2011 &	0.56\%	&  1.60\% \\
2009-2010 &	0.00\%	& -5.38\% \\
2008-2009 & 5.29\%	&  6.30\% \\
2007-2008 &	4.62\%	&  4.96\% \\
\bottomrule
\end{tabular}
\end{table}

\vskip 12pt
As can be seen, the annual rate of pay adjustment serves as an important index for the economic development of Hong Kong.
In arriving at this pay adjustment rate, a thorough consideration was often taken into account including six key factors under the established annual civil service pay adjustment mechanism.
In this study, using the annual pay adjustment, we are interested in examining whether the existing or new outlier criteria can identify the pay freeze or pay cut,
caused by the global financial crisis or the COVID-19 pandemic, as contaminated data.
Noting that the pay adjustment for the lower and mid-level staff are always the same, we thus combine these two bands and refer to them as the ``junior civil servants" for simplicity.
While for the staff in the upper band, we refer to them as the ``senior civil servants".
Lastly, for outlier detection using the three criteria, we report the pay adjustment rates since 2007 in Table \ref{HK_salary} for both the junior and senior groups, each with $n=18$ annual observations.

\vskip 6pt
For junior civil servants, the rates of pay adjustment range from 0\% to 6.16\%, together with $Q_1=2.61\%$, $Q_3=4.70\%$, and $n=18$.
The lower and upper fences of Tukey's boxplot with $k=1.5$ are ${\rm LF} = Q_1 - k\times {\rm IQR} = -0.53\%$ and ${\rm UF} = Q_3 + k\times {\rm IQR} = 7.84\%$.
Since no rate is outside the interval $[{\rm LF}, {\rm UF}]$, Tukey's boxplot fails to detect the 3 pay freezes with rate $0\%$ as contaminated data.
Next, to apply the Chauvenet-type boxplot, with $k_n^{\rm Chau} = {\Phi^{-1}(1-0.25/n)/1.35} - 0.5 = 1.13$, the lower and upper fences are given as
\begin{align*}
{\rm LF}_n^{\rm Chau} &= Q_1 - k_n^{\rm Chau}\times {\rm IQR} = 0.25\%, \\
{\rm UF}_n^{\rm Chau} &= Q_3 + k_n^{\rm Chau}\times {\rm IQR} = 7.07\%.
\end{align*}
Noting that the rate 0\% is below the lower fence, we therefore claim the 3 pay freezes in 2009-2010, 2020-2021 and 2021-2022 as contaminated data, or suspected outliers,
that were caused by either the global financial crisis or the COVID-19 pandemic.
One main reason for this success is that, due to the small sample size of $n=18$, we are able to apply a smaller fence coefficient $k_n^{\rm Chau} = 1.13$ to construct the outlier region.
Finally, it is also interesting to point out that, if Chauvenet's criterion is considered, then $\bar X = 3.51\%$, $S=2.08\%$, and $c_n = \Phi^{-1}(1-0.25/n) = 2.20$.
Further by (\ref{Chauvenet.interval}), the sigma clipping interval is given as $[\bar X - c_nS, \bar X+c_nS] = [-1.07\%, 8.09\%]$, which also fails to detect 0\% as a suspected outlier due to its sensitivity to the included outliers.

\begin{figure}
\center{
\includegraphics[width=16.5cm, angle=0]{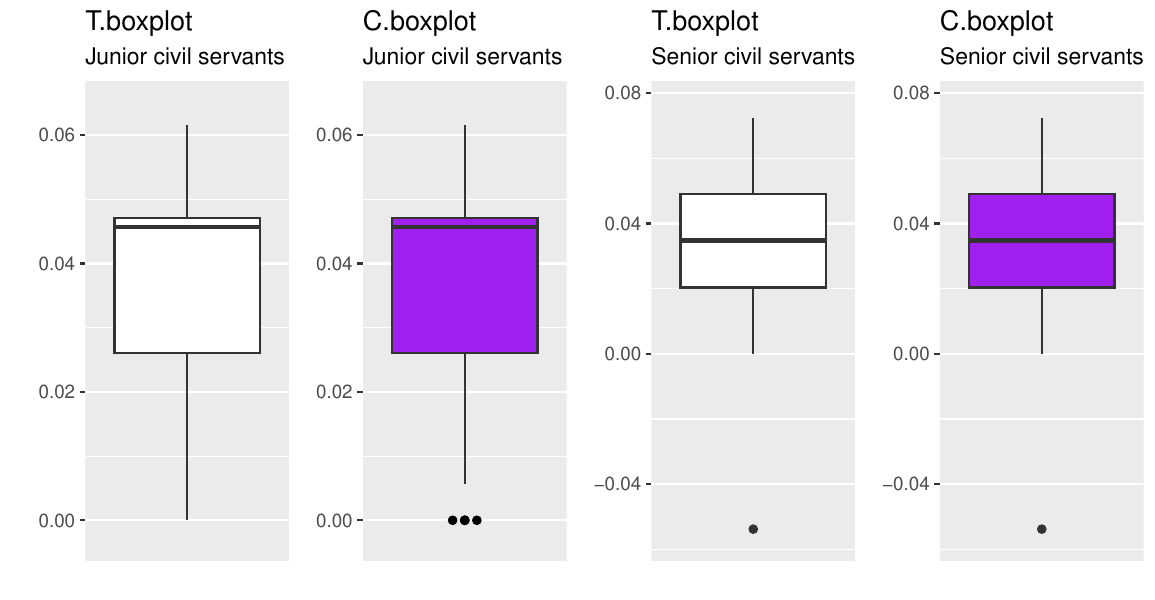}
\caption{Tukey's boxplot and the Chauvenet-type boxplot of the civil service pay adjustment in Hong Kong since 2007.
The left two panels are for junior civil servants, where the 3 outliers with jitter in the second boxplot represent the 3 pay freezes in 2009-2010, 2020-2021 and 2021-2022.
The right two panels are for senior civil servants, with the outlier in both the boxplots represents the pay cut in 2009-2010.} \label{figure5}
}
\end{figure}

\vskip 6pt
For senior civil servants, the rates of pay adjustment range from -5.38\% to 7.24\%.
In addition, we have $Q_1=2.04\%$, $Q_3=4.91\%$, $\bar X = 3.09\%$, $S=2.92\%$, and $n=18$.
For Tukey's boxplot, the lower and upper fences are ${\rm LF} = -2.27\%$ and ${\rm UF} = 9.22\%$.
For the Chauvenet-type boxplot, the lower and upper fences are ${\rm LF}_n^{\rm Chau} = -1.20\%$ and ${\rm UF}_n^{\rm Chau} = 8.15\%$.
And for Chauvenet's criterion, the sigma clipping interval is given as $[\bar X - c_nS, \bar X+c_nS] = [-3.33\%, 9.52\%]$.
It is evident that all three criteria can detect the pay cut with rate -5.38\% as an outlier of the pay adjustment data, but none of them identifies the 3 pay freezes as suspected outliers.
To clarify, with the range from -5.38\% to 7.24\%, a rate of 0\% is indeed close to the mid-range and hence does not look abnormal.
In other words, as also seen in Figure \ref{figure5}, the question of whether or not the rate 0\% is contaminated has largely been overshadowed by the abnormal rate -5.38\%.
Lastly, it is worth pointing out that our lower fence ${\rm LF}_n^{\rm Chau} = -1.20\%$ is much closer to 0\% compared to the lower fence ${\rm LF} = -2.27\%$ and the lower limit $\bar X - c_nS = -3.33\%$. This shows that, among the three criteria, our new boxplot criterion does have the greatest potential to label the pay freeze as unusual pay adjustment for the civil servants in Hong Kong.

\vskip 24pt
\section{Chauvenet-type boxplot for non-normal data}

In this section, we extend the Chauvenet-type boxplot to handle non-normal data for outlier detection.
We first show how to construct the lower and upper fences of the boxplot using the gamma distribution as an example, chosen to illustrate the method on a two-parameter distribution.
Let $\{X_1,\dots,X_n\}$ be a random sample of size $n$ from ${\rm Gamma(\alpha,\beta)}$, where $\alpha$ is the shape parameter and $\beta$ is the scale parameter \cite{Casella2002bk}.
This leads to the method of moments estimates of the two parameters as
\begin{eqnarray*}
\hat\alpha = {n\bar X^2 \over \sum_{i=1}^n (X_i - \bar X)^2} ~~~~~{\rm and}~~~~~ \hat\beta = {\sum_{i=1}^n (X_i - \bar X)^2 \over n\bar X}.
\end{eqnarray*}
We then treat ${\rm Gamma(\hat\alpha,\hat\beta)}$ as the distribution of the data, and let $\hat F^{-1}(\cdot)$ represent its quantile function.
Note that $\hat F^{-1}(\cdot)$ is a generic notation that also applies to other non-normal distributions, as long as the distribution parameters can be properly estimated from the data.

\vskip 12pt
\subsection{A new boxplot criterion for non-normal data}
Now to construct the new boxplot criterion for non-normal data, we let the lower and upper fences of the boxplot be ${\rm LF}_n = Q_1 - k_n'\times {\rm IQR}$ and ${\rm UF}_n = Q_3 + k_n''\times {\rm IQR}$, where $k_n'$ and $k_n''$ are the fence coefficients.
By the same spirit of Chauvenet's criterion, we further set $\hat F^{-1}(0.25/n)$ and $\hat F^{-1}(1-0.25/n)$ as the lower and upper thresholds for outlier detection, respectively.
In other words, Chauvenet's criterion suggests to set the lower fence as ${\rm LF}_n = \hat F^{-1}(0.25/n)$ and the upper fence as ${\rm UF}_n = \hat F^{-1}(1-0.25/n)$.
By doing so, our new criterion will wrongly reject, on average, a quarter observation from the upper tail and another quarter observation from the lower tail.
In addition, noting that $Q_1 \approx \hat F^{-1}(0.25)$ and $Q_3 \approx \hat F^{-1}(0.75)$, we can derive that
\begin{eqnarray}
k_n' \approx {\hat F^{-1}(0.25) - \hat F^{-1}(0.25/n) \over \hat F^{-1}(0.75) - \hat F^{-1}(0.25)}
~~~~~{\rm and}~~~~~
k_n'' \approx {\hat F^{-1}(1-0.25/n) - \hat F^{-1}(0.75) \over \hat F^{-1}(0.75) - \hat F^{-1}(0.25)}.   \label{Chau.coefficient.gamma}
\end{eqnarray}
In the special case when the data is normal, it is easy to verify that $k_n' = k_n'' = k_n^{\rm Chau} = \Phi^{-1}(1-0.25/n)/1.35-0.5$.
Lastly, with the fence coefficients in (\ref{Chau.coefficient.gamma}), the lower and upper fences of our new boxplot criterion can be constructed as
\begin{eqnarray}
{\rm LF}_n^{\rm Chau} = Q_1 - {\hat F^{-1}(0.25) - \hat F^{-1}(0.25/n) \over \hat F^{-1}(0.75) - \hat F^{-1}(0.25)} \times {\rm IQR},  \label{Chau.LF.gamma}
\end{eqnarray}
and
\begin{eqnarray}
{\rm UF}_n^{\rm Chau} = Q_3 + {\hat F^{-1}(1-0.25/n) - \hat F^{-1}(0.75) \over \hat F^{-1}(0.75) - \hat F^{-1}(0.25)} \times {\rm IQR}. \label{Chau.UF.gamma}
\end{eqnarray}
Lastly, to distinguish it with the Chauvenet-type boxplot in Section 3.2, we refer to the new boxplot with fences (\ref{Chau.LF.gamma}) and (\ref{Chau.UF.gamma}) as the {\it Chauvenet-type boxplot for non-normal data}.

\vskip 12pt
\subsection{Simulation study}
\begin{figure}
\center{
\includegraphics[width=16.5cm, angle=0]{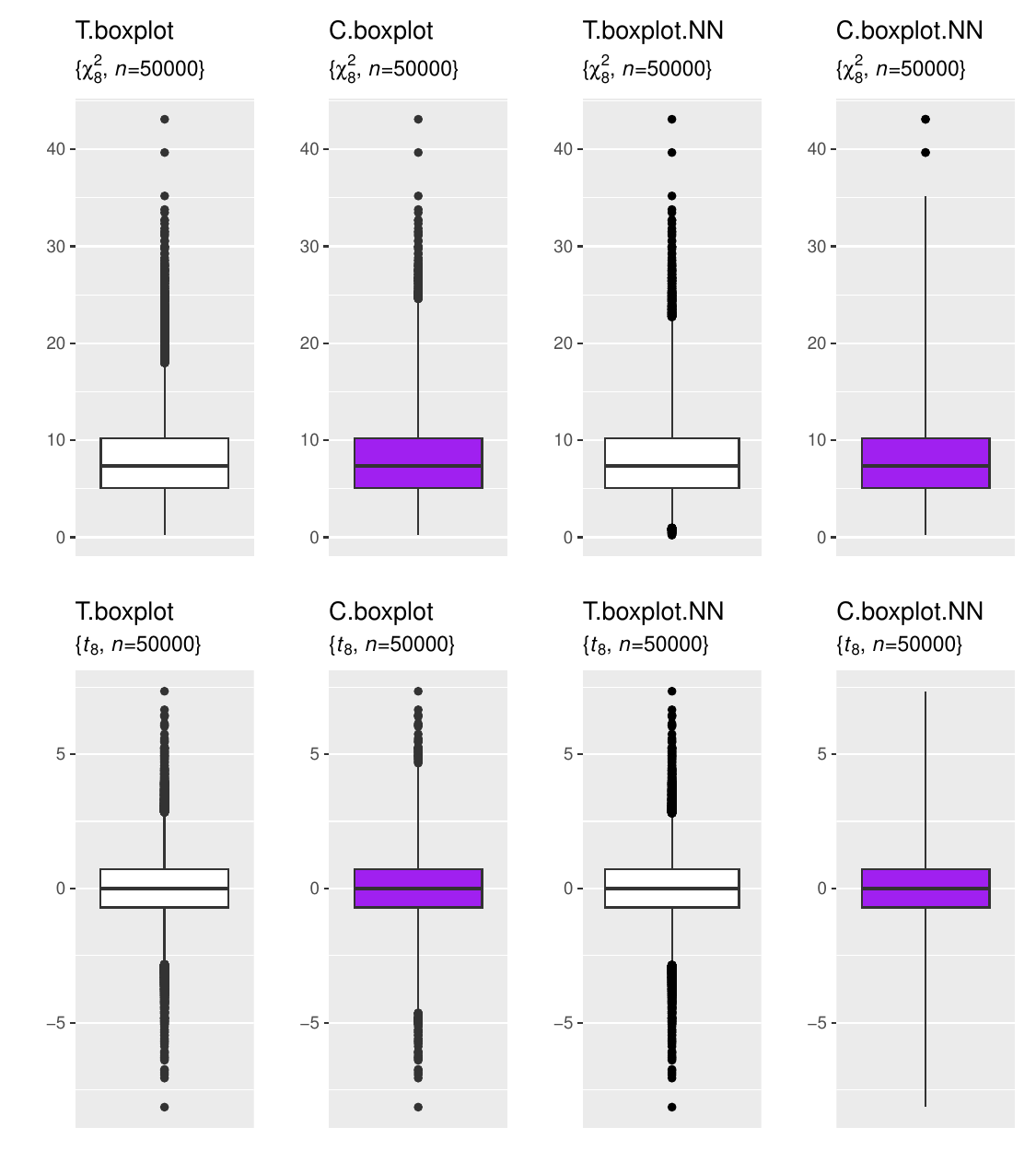}
\caption{Outlier detection for non-normal data using Tukey's boxplot (T.boxplot), the Chauvenet-type boxplot (C.boxplot), Tukey's boxplot for non-normal data (T.boxplot.NN), and the Chauvenet-type boxplot for non-normal data (C.boxplot.NN), where the data is generated from the $\chi_8^2$ and $t_8$ distributions, respectively.
With $n=50000$ and {\it set.seed(1863)} in R, the Chauvenet-type boxplot for non-normal data labels only 2 and 0 suspected outliers for each distribution.
Among the other three boxplots, the Chauvenet-type boxplot labels the least number of observations as outliers.} \label{figure6}
}
\end{figure}

To evaluate the performance of the new boxplot for non-normal data, we revisit the two distributions in Section 4.2.
We first consider the chi-square distribution with $\nu$ degrees of freedom.
Let $\{X_1,\dots,X_n\}$ be the sample data of size $n$, with $\bar X$ being the sample mean and $Q_1$ and $Q_3$ being the first and third quartiles.
Then by the method of moments, we have
\begin{eqnarray*}
\hat\nu = \bar X.
\end{eqnarray*}
Letting also $\hat F^{-1}(\cdot)$ be the quantile function of the $\chi_{\hat\nu}^2$ distribution, we can obtain the fence coefficients $k_n'$ and $k_n''$ in (\ref{Chau.coefficient.gamma}), yielding further the outlier region of the Chauvenet-type boxplot as $\left(-\infty, Q_1 - k_n'\times {\rm IQR} \right) \cup \left(Q_3 + k_n''\times {\rm IQR}, \infty \right)$.
Finally, with $\nu=8$ and $n=50000$, we generate the sample data from the $\chi_8^2$ distribution without contaminated data, and then plot the data in Figure \ref{figure6}
using Tukey's boxplot (T.boxplot), the Chauvenet-type boxplot (C.boxplot), Tukey's boxplot for non-normal data (T.boxplot.NN), and the Chauvenet-type boxplot for non-normal data (C.boxplot.NN).
More specifically for T.boxplot.NN, it is implemented using the `litteR' package in CRAN as the adjusted boxplot for skewed distributions by \citeasnoun{Hubert2008}.
Now if we take {\it set.seed(1863)} in R, the simulated data will range from $0.22$ to $43.09$, with $Q_1=5.08$, $Q_3=10.24$ and $\hat\nu = \bar X =8.02$.
Further by (\ref{Chau.coefficient.gamma})-(\ref{Chau.UF.gamma}), it leads to $k_n' = 0.94$, $k_n'' = 5.58$, and $[{\rm LF}_n^{\rm Chau}, {\rm UF}_n^{\rm Chau}] = [0.20, 39.02]$.
Lastly, observing that there are only two observations (39.66 and 43.09) above the upper fence and no observation below the lower fence, C.boxplot.NN thus only labels two suspected outliers among a total of 50000 observations.
Among the other three boxplots, however, the Chauvenet-type boxplot performs the best since it only labels 106 suspected outliers as already seen in Section 4.2.

\vskip 6pt
For the second simulation, we consider Student's $t$ distribution with $\nu$ degrees of freedom.
Let $\{X_1,\dots,X_n\}$ be the sample data with $S^2$ being the sample variance and $Q_1$ and $Q_3$ being the first and third quartiles.
By equating the sample variance to the population variance, i.e. $\nu/(\nu-2) = S^2$, we have the estimated degrees of freedom as
\begin{eqnarray*}
\hat\nu = {2S^2 /(S^2-1)}.
\end{eqnarray*}
Letting also $\hat F^{-1}(\cdot)$ be the quantile function of the $t_{\hat\nu}$ distribution, we can thus have the outlier region of the Chauvenet-type boxplot as $\left(-\infty, Q_1 - k_n'\times {\rm IQR} \right) \cup \left(Q_3 + k_n''\times {\rm IQR}, \infty \right)$.
Finally, with $\nu=8$ and $n=50000$, we generate data from the $t_8$ distribution and also plot the four boxplots in Figure \ref{figure6} as in the previous simulation.
Taking {\it set.seed(1863)} in R again, the simulated data ranges from $-8.13$ to $7.34$, with $Q_1=-0.70$, $Q_3=0.71$, $\hat\nu = {2S^2 /(S^2-1)} = 8.02$ and $k_n' = k_n'' = 6.41$.
This leads to $[{\rm LF}_n^{\rm Chau}, {\rm UF}_n^{\rm Chau}] = [-9.77, 9.78]$, thus explaining why C.boxplot.NN labels no outlier.
Lastly, we note that the comparison results among the other three boxplots remain the same.
For additional results on the boxplots with $n=50$, 500 and 5000 from the $\chi_8^2$ and $t_8$ distributions, please refer to the online Supplementary Material.

\vskip 24pt
\section{Conclusion}

Tukey's box-and-whisker plot is one of the most popular methods used for descriptive statistics, but it is also known to be free of sample size, yielding the so-called ``one-size-fits-all" fences for outlier detection.
The main purpose of this paper was to introduce a new boxplot criterion that can more accurately, or more robustly, label the suspected outliers or contaminated data.
In contrast to Tukey's boxplot with the whiskers being $1.5$ or $3$ times of the interquartile range (${\rm IQR}$), our new boxplot tailored the whiskers' length as $k_n^{\rm Chau}\times {\rm IQR}$, where
\begin{eqnarray*}
k_n^{\rm Chau} = {\Phi^{-1}(1-0.25/n) \over 1.35} - 0.5
\end{eqnarray*}
is the sample size adjusted fence coefficient determined by Chauvenet's criterion, with $n$ being the sample size and $\Phi^{-1}$ being the quantile function of the standard normal distribution.
Observing that our new boxplot takes advantage of both Tukey's boxplot and Chauvenet's criterion, we referred to it as the {\it Chauvenet-type boxplot},
with the lower and upper fences being specified as ${\rm LF}_n^{\rm Chau} = Q_1 - k_n^{\rm Chau}\times {\rm IQR}$ and ${\rm UF}_n^{\rm Chau} = Q_3 + k_n^{\rm Chau}\times {\rm IQR}$ as in (\ref{Chau.fence}).
In addition, we noted that $k_n^{\rm Chau}$ is an increasing function of $n$, with $k_n^{\rm Chau}=1.5$ when $n=72$, and $k_n^{\rm Chau}=3$ when $n=217,282$.
And in particular, when the sample size is less than $72$, we even recommended a whisker shorter than $1.5\times {\rm IQR}$ for outlier detection, in a way to more effectively increase the detection power of outliers.
With the fence coefficient $k_n^{\rm Chau}$, the Chauvenet-type boxplot not only provides an exact control of the outside rate per observation, but also maintains the simplicity of the boxplot from a practical perspective.
Moreover, from the perspective of Chauvenet's criterion, our new boxplot criterion can also serve as a robust Chauvenet's criterion.
Simulation study and a real data analysis on the civil service pay adjustment in Hong Kong also demonstrated that the Chauvenet-type boxplot performs extremely well regardless of the sample size, and can therefore be highly recommended for practical use to replace both Tukey's boxplot and Chauvenet's criterion.

\vskip 6pt
Lastly, we also presented in Section 5 that the Chauvenet-type boxplot can be further extended to handle non-normal data for outlier detection.
By numerical studies with the data generated from the chi-square and $t$ distributions, the {\it Chauvenet-type boxplot for non-normal data} can perfectly do its job,
and in particular by providing two different whiskers, our new criterion also effectively prevents to label too many outliers from one side of the skewed data.
On the other hand, we noted that the fence coefficients $k_n'$ and $k_n''$ derived in (\ref{Chau.coefficient.gamma}) require the non-normal distribution to be known, which however may not be realistic in practice.
If the underlying distribution of the sample data is unknown, one may need to apply other statistical techniques, e.g. the nonparametric methods, to estimate the distribution function.
As another alternative, one may also apply some existing techniques in the literature to label outliers for non-normal and in particular for skewed data
\cite{Schwertman2004,Dumbgen2007,Schwertman2007,Hubert2008,Bruffaerts2014,Dovoedo2015,Shein2017,Walker2018,ZhaoC2019,Rodu2022}.
In particular, we noted that \citeasnoun{Kimber1990} applied the semi-interquartile ranges to construct the lower and upper fences, yielding ${\rm LF} = Q_1-k(M-Q_1)$ and ${\rm UF} =  Q_3+k(Q_3-M)$,
and \citeasnoun{Carling2000} further constructed the lower and upper fences as a form of ${\rm LF} = M - k\times {\rm IQR}$ and ${\rm UF} = M + k\times {\rm IQR}$, where $M$ is the sample median and $k$ is the fence coefficient.
Further research is warranted in this direction to build more robust Chauvenet-type boxplots.

\vskip 24pt
\section*{Supplementary Material}
The online Supplementary Material gives the proof of Theorem 1 together with a supporting lemma from the literature.
It also provides a total of nine additional figures for non-normal boxplots with other skewed or heavy tailed distributions to supplement the simulation studies in the main text.

\vskip 24pt
\section*{Acknowledgments}
The authors thank the editor, the associate editor, and the two reviewers for their constructive comments that have led to a significant improvement of the paper, in particular on the quality of the figures and the user-friendly R package `ChauBoxplot' for implementing the Chauvenet-type boxplot.
Hongmei Lin's research was supported in part by the National Natural Science Foundation of China (12171310), the Shanghai ``Project Dawn 2022" (22SG52), and the Basic Research Project of Shanghai Science and Technology Commission (22JC1400800).
Riquan Zhang's research was supported in part by the National Natural Science Foundation of China (12371272) and the Basic Research Project of Shanghai Science and Technology Commission (22JC1400800).
Tiejun Tong's research was supported in part by the General Research Fund of Hong Kong (HKBU12300123 and HKBU12303421) and the Initiation Grant for Faculty Niche Research Areas of Hong Kong Baptist University (RC-FNRA-IG/23-24/SCI/03).

\vskip 24pt
\section*{Disclosure Statement}
The authors report there are no competing interests to declare.

\vskip 24pt
\bibliographystyle{dcu}
\citationstyle{dcu}
\bibliography{references_18Feb2024}

\end{document}